\documentclass[prl,twocolumn,showpacs,superscriptaddress]{revtex4}

\usepackage{amsfonts,amssymb,amsmath,amsthm,epsfig} 
\usepackage{mathrsfs}
\usepackage{graphicx}
\usepackage{array}

\newcommand{\Ga}{\Gamma}
\newcommand{\ka}{\kappa}
\newcommand{\De}{\Delta}
\newcommand{\de}{\delta}

\newcommand{\beqs}{\begin{equation*}}
\newcommand{\beq}{\begin{equation}}

\newcommand{\eeqs}{\end{equation*}}
\newcommand{\eeq}{\end{equation}}

\newcommand{\beqas}{\begin{eqnarray*}}
\newcommand{\beqa}{\begin{eqnarray}}

\newcommand{\eeqas}{\end{eqnarray*}}
\newcommand{\eeqa}{\end{eqnarray}}

\newcommand{\eq}[2]{\begin{equation} #1 \label{#2} \end{equation}}

\newcommand{\blist}{\begin{itemize}}

\newcommand{\elist}{\end{itemize}}

\providecommand{\href}[2]{#2}

\newcommand{\pont}{{\,^\ast\!}R\,R}

\begin{document}

\title{Dirichlet boundary value problem for Chern-Simons modified gravity}

\author{Daniel Grumiller}
\affiliation{Center for Theoretical Physics,
  Massachusetts Institute of Technology, 77 Massachusetts Ave., Cambridge, MA 02139, USA}
\email{grumil@lns.mit.edu}

\author{Robert Mann}
\affiliation{Department of Physics \& Astronomy, University of Waterloo,
Waterloo, Ontario N2L 3G1, Canada}
\email{rbmann@sciborg.uwaterloo.ca}
\affiliation{Perimeter Institute, 31 Caroline Street North, Waterloo, Ontario N2L 2Y5, Canada}
\email{rmcnees@perimeterinstitute.ca}

\author{Robert McNees}
\affiliation{Perimeter Institute, 31 Caroline Street North, Waterloo, Ontario N2L 2Y5, Canada}
\email{rmcnees@perimeterinstitute.ca}

\date{\today}

\preprint{MIT-CTP 3933}

\begin{abstract}

Chern-Simons modified gravity comprises the Einstein-Hilbert action and a higher-derivative interaction containing the Chern-Pontryagin density. We derive the analog of the Gibbons-Hawking-York boundary term required to render the Dirichlet boundary value problem well-defined. It turns out to be a boundary Chern-Simons action for the extrinsic curvature. We address applications to black hole thermodynamics.

\end{abstract}


\maketitle

In 1744-1746 Maupertuis formulated a General Principle \cite{Maupertuis:1746}: {\em When a change occurs in Nature, the quantity of action necessary for that change is as small as possible.} In other words, the dynamics of a system follow from the condition that the first variation of the action vanishes. This includes all total derivative terms, which are addressed by imposing appropriate boundary conditions on the fields. Thus, a well-defined boundary value problem is a prerequisite for any new theory formulated by means of an action principle.

Theories of gravity are notorious in this regard: The Einstein-Hilbert action of general relativity,
\eq{
S_{EH}\sim\int d^4x\,\sqrt{-g}\; R
}{eq:EH}
does not admit a well-defined Dirichlet boundary value problem unless it is supplemented by the Gibbons-Hawking-York (GHY) boundary term \cite{York:1972sj,Gibbons:1976ue}. This is because the Ricci scalar $R$ contains second derivatives of the metric, so variation of the action produces boundary terms involving the variation of both the metric and its normal derivative. The GHY term cancels the latter contributions, and the remaining terms are addressed by placing Dirichlet boundary conditions on the metric.

String theory and other UV-completions of Einstein gravity suggest that higher powers of curvature invariants appear in the low energy effective action, suppressed by powers of the Planck-mass. These actions can also be considered as models in their own right, but in that case the problem described above reappears.
Higher powers of the curvature introduce more derivatives of the metric, and the Dirichlet boundary value problem is no longer well-defined. A notable exception is Lovelock gravity \cite{Lovelock:1971yv}, where it is possible to render the Dirichlet boundary value problem well-defined by adding appropriate boundary terms that are non-linear in extrinsic curvature \cite{Myers:1987yn}.

Recently a CP-odd modification of general relativity dubbed ``Chern-Simons modified gravity'' was proposed. The action for this theory takes the form $S=S_{EH}+S_{CS}$ ~\cite{Jackiw:2003pm}, where
the new term is given by
\begin{equation}
\label{eq:intro}
S_{CS} \sim \int d^4x\,\sqrt{-g}  \;  \theta \, \pont \,.
\end{equation}
The action will be described in more detail momentarily.
Here we just mention that $\pont$ is the Chern-Pontryagin term and $\theta$ is a background scalar field, similar to an axion. The modification \eqref{eq:intro} has engendered a lot of interesting research --- cf.~e.g.~\cite{Jackiw:2003pm,Kostelecky:2003fs,
Grumiller:2007rv,Weinberg:2008hq} and Refs.~therein --- but so far the boundary value problem has not been addressed. Is there a boundary term analogous to the GHY term that can be added to the action to give a well-defined Dirichlet boundary value problem?

We answer this question in the affirmative by constructing the boundary term required by the Chern-Simons (CS) modification \eqref{eq:intro}. Schematically, our main result is that \eqref{eq:intro} has to be supplemented by a boundary term quadratic in the extrinsic curvature
\eq{
S_{bCS} \sim \int d^3x\,\sqrt{h}\;\theta\,{^{\,\ast\!}K}DK
\sim \int d^3x \,\sqrt{h}\;\theta \, {\rm CS}(K) \,.
}{eq:mainresult}
This term may be interpreted as a boundary CS action for the (traceless part of) extrinsic curvature.

Our conventions are as follows. We use signature $-,+,+,+$. Indices from the beginning of the alphabet $a,b,\dots$ range from $0$ to $3$ while indices from the middle of the alphabet $i,j,\dots$ range from $0$ to $2$. Conventional factors and signs are chosen so that $A_{(ab)}:=(A_{ab}+A_{ba})/2$, the Riemann tensor has sign $R^a{}_{bcd}:= \partial_c\Ga^a{}_{bd} - \dots$, and $R_{ab}:=R^c{}_{acb}$. We denote the $\epsilon$-tensor by $\epsilon^{abcd}$ and $\epsilon^{ijk}$, and the $\varepsilon$-symbol by $\varepsilon^{abcd}$ and $\varepsilon^{ijk}$. Covariant derivatives in four (three) dimensions are denoted by $\nabla_a$ ($D_i$), partial derivatives by $\partial_a$ ($\partial_i$).

First we recall the basics of CS modified gravity. The bulk part of the action is given by~\cite{Jackiw:2003pm}\footnote{Adding kinetic and potential terms for the scalar field $\theta$ is possible, but does not change anything essential about boundary issues as long as $\theta$ obeys appropriate boundary conditions.}
\beq
\label{CSaction}
S = S_{EH}+S_{CS}=\kappa \int d^{4}x\, \sqrt{-g}\, \Big(R + \frac{1}{4} \,\theta \, \pont
\Big) \,,
\eeq
where $\kappa$ is the gravitational coupling constant
and $\pont$ is the Chern-Pontryagin term, defined as
\beq
\label{pontryagindef}
\pont:={\,^\ast\!}R^a{}_b{}^{cd} R^b{}_{acd}=\frac12 \epsilon^{cdef}R^a{}_{bef} R^b{}_{acd} ~.
\eeq
Equation \eqref{pontryagindef} can be expressed as the covariant divergence
\beq
\nabla_a J^a = \frac12 \pont
\label{eq:curr1}
\eeq
of the CS topological current ($\Gamma^a{}_{bc}$ are the Christoffels),
\beq
J^a :=\epsilon^{abcd}\Big(\Gamma^n{}_{bm}\partial_c\Gamma^m{}_{dn}+\frac23\Gamma^n{}_{bm}\Gamma^m{}_{cl}\Gamma^l{}_{dn}\Big)\,,
\label{eq:curr2}
\eeq
hence the name ``Chern-Simons modified gravity''.

The modified field equations are obtained by varying the action
with respect to the metric. Using
\beq
\delta R^b{}_{acd}=\nabla_c\delta\Gamma^b{}_{ad}-\nabla_d\delta\Gamma^b{}_{ac}
\label{eq:deR}
\eeq
and
\beq
\delta\Gamma^b{}_{ac} = \frac12 g^{bd}\left(\nabla_a\delta
  g_{dc}+\nabla_c\delta g_{ad}-\nabla_d\delta g_{ac}\right)\,,
\label{eq:deGa}
\eeq
the variation of the action is ($G_{ab}=R_{ab}-\frac12 g_{ab}R$)
\beqa
\label{variationofS}
\delta S &=& -\kappa \int d^4x\, \sqrt{-g}\,
  \left(G^{ab} + C^{ab} \right) \delta g_{ab}
\nonumber \\
&+& {\rm boundary\;terms}\,.
\eeqa
The symmetric, traceless tensor $C^{ab}$ is given by
\beq
\label{Ctensor}
C^{ab} := (\nabla_c\theta)\,
\epsilon^{cde(a}\nabla_eR^{b)}{}_d+(\nabla_{(c}\nabla_{d)}\theta){\,^\ast\!}R^{d(ab)c}\,.
\eeq
Surface terms that arise from repeated integration by
parts are collected in the second line of
Eq.~(\ref{variationofS}). They will be studied in detail below.

Provided the boundary terms are dealt with appropriately, the modified
field equations can be written as
\beq
\label{eom}
R_{ab} + C_{ab} = 0\,.
\eeq
If $\theta$ is not a constant then the contracted Bianchi identities applied to \eqref{eom} imply the
so-called Pontryagin constraint
\eq{
\pont = 0 \,.
}{eq:pont}
The main qualitative difference to general relativity is the emergence of
first derivatives of the Ricci tensor or, equivalently, of third derivatives of the metric in the equations of motion.
The appearance of these higher derivatives has important repercussions on
boundary issues.

Turning to the second line of Eq.~\eqref{variationofS},  we are specifically interested in boundary terms involving normal derivatives of the metric variation. This is a small subset of the terms that appear when varying quantities like the action or extrinsic curvature, so we use a special notation to isolate them. Equivalence between two quantities up to `irrelevant terms' is denoted by $\simeq$. By definition `irrelevant terms' are bulk terms that are not total derivatives, or boundary terms that vanish when Dirichlet boundary conditions are imposed on the metric.

Before dealing with the modified action \eqref{CSaction} we review the origin of the GHY boundary term. The variation of the Einstein-Hilbert action is
\begin{align}
\de S_{EH} & = \ka \int d^4x\,\sqrt{-g}\, ( g^{ab}\de R_{ab} -G^{ab} \de g_{ab})\nonumber \\
 & \simeq \ka \int d^4x\,\sqrt{-g}\;g^{ab}\Big(\nabla_c\de\Ga^c{}_{ab}-\nabla_b\de\Ga^c{}_{ac}\Big) ~.
\label{eq:bCS1}
\end{align}
We assume from now on that the boundary is a hypersurface with spacelike outward-pointing unit vector $n^a$, i.e., $n^a n_a=1$ \,\footnote{For simplicity we ignore past and future components of the boundary. It is straightforward to adapt our results to components of the boundary with a timelike normal.}. Then the induced metric on the boundary is given by
\eq{
h_{ab} = g_{ab}-n_a n_b ~.
}{eq:bCS2}
The extrinsic curvature is the Lie derivative of $\frac12 h_{ab}$ along $n^{a}$
\eq{
K_{ab}=\frac12 {\cal L}_n h_{ab} = h_a^c h_b^d \nabla_c n_d\,,
}{eq:bCS3}
with trace
$K=K^a{}_a=\nabla_a n^a$.
The term in the variation of the extrinsic curvature relevant to our calculations is given by
\eq{
\delta K_{ab}
\simeq \frac12 h_a^c h_b^d n^e \nabla_e \delta g_{cd}\,.
}{eq:bCS5}
Applying Eqs.~\eqref{eq:deGa} and \eqref{eq:bCS2} to Eq.~\eqref{eq:bCS1} and comparing with Eq.~\eqref{eq:bCS5} yields
\eq{
\delta S_{EH} \simeq - 2 \kappa \,\delta  \int d^{3}x \,\sqrt{h}\; K \,.
}{eq:angelinajolie}
This leads to the familiar result that the action
\eq{
S_{EH}+S_{GHY} = \kappa \int d^4x \,\sqrt{-g}\; R + 2\ka \int d^3x \,\sqrt{h}\; K
}{eq:bCS7}
has a well-defined Dirichlet boundary problem \cite{York:1972sj,Gibbons:1976ue}.

We now perform a similar analysis for the second term in Eq.~\eqref{CSaction}. Applying Eq.~\eqref{eq:deR} to Eq.~\eqref{CSaction} yields
\eq{
\delta S_{CS}\simeq - \ka\int d^4x\,\sqrt{-g}\;\nabla_c\left(\theta ^{\,\ast\!} R_a{}^{bcd}\de\Ga^a{}_{bd}\right)\,.
}{eq:bCS8}
To proceed we need a 3+1 decomposition of all tensors with respect to the induced metric and the normal vector. For simplicity we introduce an adapted coordinate system where the shift vector vanishes and the lapse function is unity, but the final result will be given in a manifestly covariant form. We denote tangential indices by $i,j,\dots$ and indices contracted with the normal vector by ${\bf n}$.
With respect to this decomposition we obtain $^{\,\ast\!}R_a{}^{b{\bf n}d}\de\Ga^a{}_{bd} = {^{\ast\!}R}_j{}^{k{\bf n}i}\de\Ga^j{}_{ki}+2^{\,\ast\!}R_{\bf n}{}^{j{\bf n}i}\de\Ga^{\bf n}{}_{ji}\simeq 2^{\,\ast\!}R_{\bf n}{}^{j {\bf n}i}\de\Ga^{\bf n}{}_{ji}$ and Eq.~\eqref{eq:bCS8} reduces to
\eq{
\delta S_{CS}\simeq -2\ka\int d^4x\,\sqrt{-g}\;\nabla_{\bf n}\left(\theta ^{\,\ast\!} R_{\bf n}{}^{i{\bf n}j}\de\Ga^{\bf n}{}_{ij}\right)\,.
}{eq:bCS9}
By virtue of Eq.~\eqref{eq:deGa} this further simplifies to
\eq{
\delta S_{CS}
\simeq -\ka\int d^3x\,\sqrt{h}\;\theta ^{\,\ast\!} R^{i{\bf n n}j}\nabla_{\bf n}\de g_{ij}
}{eq:bCS10}
The decomposition of the dual Riemann tensor
$^{\,\ast\!} R^{i{\bf n}{\bf n}j}=\frac12 \epsilon^{{\bf n}jkl}R^{i{\bf n}}{}_{kl}$ requires us to calculate $R^{i{\bf n}}{}_{kl}$. Let $D_i$ be the covariant derivative along the boundary whose connection $\gamma^i{}_{jk}$ is torsionfree and compatible with the induced metric $h_{ij}$. Then the Codazzi equation
$R_{{\bf n}ijk} = D_k K_{ij}-D_j K_{ik}$
together with the variation Eq.~\eqref{eq:bCS5} can be used to rewrite Eq.~\eqref{eq:bCS10} as
\eq{
\delta S_{CS}\simeq  2\ka \int d^3x\,\sqrt{h} \;\theta \,\epsilon^{ijk} \, (D_j \,K_i{}^l)\,\delta K_{kl}\,,
}{eq:bCS12}
where $\epsilon^{ijk} = \epsilon^{{\bf n}ijk} $.
The expression \eqref{eq:bCS12} can be cancelled by adding the following boundary term to the action
\eq{
S_{bCS} =\ka \int d^3x\,\sqrt{h} \;\theta \,\epsilon^{ijk} \,K_i{}^{l} \,D_j \,K_{kl}\,,
}{eq:bCS12a}
i.e., $\de S_{CS}\simeq-\de S_{bCS}$.
With the abbreviation
\eq{
{\rm CS}(K):= {^{\,\ast\!}K}^{jkl}D_j K_{kl} =
\frac12 \,\epsilon^{ijk} \,K_i{}^l \,D_j \,K_{kl}
}{eq:bCS13}
we obtain the result announced in Eq.~\eqref{eq:mainresult}. The new boundary term depends only on the traceless part of the extrinsic curvature, and is therefore complementary to the GHY term in \eqref{eq:bCS7}. The result \eqref{eq:bCS12a} might have been anticipated on the grounds that it is required by the index theorem for manifolds with boundary, cf.~e.g.~section 8 in Ref.~\cite{Eguchi:1980jx}.

The abbreviation ``CS'' emphasizes the fact that \eqref{eq:bCS13} resembles an abelian Chern-Simons term.
It should be contrasted with the gravitational CS term in three dimensions \cite{Deser:1982vy}
\eq{
{\rm CS}(\gamma)=\frac12 \epsilon^{ijk} \Big(\gamma^l{}_{im}\partial_j\gamma^m{}_{kl}+\frac23\gamma^l{}_{im}\gamma^m{}_{jp}\gamma^p{}_{kl}\Big)\,,
}{eq:bCSnew}
constructed from the intrinsic connection $\gamma^i{}_{jk}$. It is interesting to note that the terms comprising the CS modification are equivalent to
\begin{multline}
\frac{1}{4} \int d^4x \,\sqrt{-g} \;\theta \, \pont + 2 \int d^3x\, \sqrt{h} \; \theta \,{\rm CS}(K) \\
= -\frac12 \int d^4x\,\sqrt{-g}\,(\nabla_a\theta) J^a+\int d^3x\,\sqrt{h}\; \theta\,{\rm CS}(\gamma)\,,
\label{eq:conclusion}
\end{multline}
where $J^{a}$ is the topological current defined in Eq. \eqref{eq:curr2}.

We conclude that the full action for CS modified gravity is given by
\begin{multline}
S = S_{EH}+S_{CS}+S_{GHY}+S_{bCS} = \\
=\kappa \int d^4x \,\sqrt{-g} \,\Big(R + \frac{1}{4} \,\theta \, \pont \Big) \\ + 2\kappa \int d^3x\, \sqrt{h} \,\Big(K + \theta \,{\rm CS}(K)\Big) ~.
\label{eq:fullaction}
\end{multline}
The action \eqref{eq:fullaction} has a well-defined Dirichlet boundary value problem. This is our main result.
We present now Eq.~\eqref{eq:fullaction} in a manifestly covariant form:
\begin{align}
S
&= \kappa \int d^4x \,\sqrt{-g} \,\Big(R + \frac{1}{4} \,\theta \, \pont \Big) \nonumber \\
& + 2\kappa \int d^3x\, \sqrt{h}\, \Big(K + \frac12 \theta \,n_a\epsilon^{abcd}K_b{}^e\nabla_c K_{de}\Big) \nonumber \\
& \;+\ka \int d^3x\,\sqrt{h}\;{\cal F}(h_{ab},\theta) ~.
\label{eq:covariantaction}
\end{align}
The last line is an additional term $S_{{\cal F}}$ that is intrinsic to the boundary.
This `boundary counterterm' does not affect the Dirichlet boundary value problem, and is necessary for a well-defined variational principle when the boundary is removed to spatial infinity \cite{Mann:2005yr, Mann:2006bd}.

The boundary terms in Eq.~\eqref{eq:covariantaction} are required for self-consistency of the theory and a prerequisite to a Hamiltonian formulation. They also contribute to the semi-classical approximation of the thermodynamical partition function ${\cal Z}$ in the Euclidean path integral approach \cite{Gibbons:1994cg},~\footnote{The Einstein-Hilbert action $S_{EH}$ does not contribute to the on-shell action because the modified field Eqs.~\eqref{eom} imply $R=0$.}:
\eq{
{\cal Z} \approx
\exp{[-(S_{CS}+S_{GHY}+S_{bCS}+S_{\cal F})|_{\rm on-shell}]}
}{eq:bCS16}
A prime application where these boundary terms are important is black hole thermodynamics. To investigate this issue we make the ansatz
\eq{
{\cal F}={\cal F}_0 (h_{ab}) + \De {\cal F} (h_{ab},\theta)\,,
}{eq:bCS17}
where ${\cal F}_0$ is the boundary counterterm required by general relativity and $\De {\cal F}$ is a new contribution that must be linear in $\theta$~\footnote{
The linearity in $\theta$ can be seen as follows. The new contribution to the bulk action and the novel boundary term constructed in the present work are both linear in $\theta$. Therefore, all new contributions spoiling the variational principle must be linear in $\theta$ as well, and consequently the boundary counterterm which repairs this deficiency must also be linear in $\theta$.
}. The contributions to the on-shell action that differ from general relativity are $\De S=S_{CS}+S_{bCS}+S_{\De{\cal F}}$:
\begin{multline}
\De S|_{\rm on-shell} = \frac{\kappa}{4} \int d^4x \, \sqrt{-g} \;\theta \, \pont \\
+\kappa \int d^3x\, \sqrt{h}\,\Big(2\,\theta\,{\rm CS}(K) + \De {\cal F}(h_{ab},\theta)\Big)
\label{eq:bCS18}
\end{multline}
The bulk term in Eq.~\eqref{eq:bCS18} vanishes due to the constraint \eqref{eq:pont}. Therefore, all new contributions to the on-shell action are due to the boundary terms in the second line.

An interesting subtlety occurs when $\theta=\theta_0$ is a constant. In that case the field equations reduce to the Einstein equations, and the constraint \eqref{eq:pont} no longer applies. However, the on-shell action still receives a contribution from the CS modification,
\eq{
\De S|_{\rm on-shell} = -4\pi^2\ka\,\theta_0\,\tau({\cal M}) + \ka\,\theta_0 \int d^3x\,\sqrt{h}\;\De {\cal F}(h_{ab})\,,
}{eq:bCS19}
where
\begin{multline}
\tau({\cal M})=\frac{1}{32\pi^2} \int_{\cal M} \!\!d^4x\; \varepsilon^{cdef} R^a{}_{bef} R^b{}_{acd} \\
- \frac{1}{4\pi^2} \int_{\partial{\cal M}} \!\!\!\!d^3x\; \varepsilon^{ijk} K_i{}^l D_j K_{kl}
\label{eq:bCS20}
\end{multline}
is the signature index for a manifold ${\cal M}$ with boundary $\partial {\cal M}$ (for a list of examples with $\tau\neq 0$ see table D.1 in Ref.~\cite{Eguchi:1980jx}). Since $\tau({\cal M})$ is just some (integer) number, it does not modify black hole thermodynamics.

The result \eqref{eq:bCS18} holds in general, but there may be additional simplifications for specific black hole solutions.
For instance, the Schwarzschild solution persists in CS modified gravity \cite{Jackiw:2003pm} (at least for certain choices of the scalar field $\theta$ \cite{Grumiller:2007rv}), because the two tensors $R_{ab}$ and $C_{ab}$ vanish separately and thus the field Eqs.~\eqref{eom} are fulfilled. We consider now the Schwarzschild black hole in CS modified gravity and assume that the boundary is a hypersurface of constant surface area. Because it is possible to simultaneously  diagonalize the metric and the extrinsic curvature, the expression CS$(K)$ defined in Eq.~\eqref{eq:bCS13} vanishes identically. Thus, the only remaining new contribution to the on-shell action comes from $\De {\cal F}$. We conclude that the thermodynamics of the Schwarzschild black hole in CS modified gravity is the same as the thermodynamics of the Schwarzschild black hole in general relativity, up to possible corrections from the modification $\De {\cal F}$ of the boundary counterterm.

Another interesting application of our analysis is that it provides a short-cut to an effective field theory of single field inflation. Namely, after writing down all terms with four derivatives one obtains a correction to the action containing ten terms (see Eq.~(3) in Ref.~\cite{Weinberg:2008hq}). Requiring a well-defined Dirichlet boundary value problem eliminates seven of these terms. The remaining three terms are identical to the ones contained in Eq.~(6) of Ref.~\cite{Weinberg:2008hq} and include the Gauss-Bonnet term studied in Ref.~\cite{Myers:1987yn} and the gravitational Chern-Simons term studied in the present work. 

In this Letter we constructed the analog of the Gibbons-Hawking-York term for Chern-Simons modified gravity and showed that the action Eq.~\eqref{eq:covariantaction} has a
well-defined Dirichlet boundary value problem. However, for physically
interesting boundary conditions this does not guarantee a
well-defined variational problem. Even asymptotically flat
boundary conditions require additional surface terms in the action \cite{Mann:2005yr, Mann:2006bd}.
The structure of these terms is currently being investigated and will be reported in a future
publication.

\acknowledgments

We thank Roman Jackiw and Nicolas Yunes for discussions.
This work is supported in part by funds provided by the U.S. Department of Energy (D.O.E.) under the cooperative research agreement DEFG02-05ER41360 and by the Natural Sciences and Engineering
Research Council of Canada. DG is supported by the project MC-OIF 021421 of the European Commission under 
FP6. RM$^2$ are supported by the Perimeter Institute for Theoretical Physics (PI). Research at PI is supported in part by funds from NSERC of Canada and MEDT of Ontario. DG acknowledges travel support by PI and the kind hospitality at PI while part of this Letter was conceived.


\end{document}